\documentclass[conference]{IEEEtran}
\input{psfig.sty}
\usepackage{subfigure}
\usepackage{multirow}
\usepackage{psfig}
\usepackage{graphicx}
\usepackage{amsmath}
\usepackage{color}
\usepackage{epsfig}
\usepackage{amsfonts}
\usepackage{algorithm2e}
\usepackage{subfigure}

\newcommand{\Define}{\stackrel{\triangle}{=}}

\begin{document}
\twocolumn

\title{{\LARGE
$M$-ary Detection and $q$-ary Decoding in Large-Scale MIMO: A Non-Binary 
Belief Propagation Approach
}}
\author{T. Lakshmi Narasimhan and A.~Chockalingam \\
{\normalsize Department of ECE, Indian Institute of Science,
Bangalore 560012, India } }

\IEEEaftertitletext{\vspace{-0.6\baselineskip}}
\maketitle
\begin{abstract}
In this paper, we propose a non-binary belief propagation approach (NB-BP) for 
detection of $M$-ary modulation symbols and decoding of $q$-ary LDPC codes in 
large-scale multiuser MIMO systems. We first propose a message passing based 
symbol detection algorithm which computes vector messages using a scalar 
Gaussian approximation of interference, which results in a total complexity 
of just $O(KN\sqrt{M})$, where $K$ is the number of uplink users and $N$ is 
the number of base station (BS) antennas. The proposed NB-BP detector 
does not need to do a matrix inversion, which gives a complexity advantage
over MMSE detection. We then design optimized $q$-ary LDPC 
codes by matching the EXIT charts of the proposed detector and the LDPC decoder. 
Simulation results show that the proposed NB-BP detection-decoding approach 
using the optimized LDPC codes achieve significantly better performance (by about 
1 dB to 7 dB at $10^{-5}$ coded BER for various system loading factors with number 
of users ranging from 16 to 128 and number of BS antennas fixed at 128) compared 
to using linear detectors (e.g., MMSE detector) and off-the-shelf $q$-ary irregular 
LDPC codes. Also, even with estimated channel knowledge (e.g., with MMSE channel 
estimate), the performance of the proposed NB-BP detector is better than that 
of the MMSE detector.
\end{abstract}

\section{Introduction}
\label{sec1} 
Multiple-input multiple-output (MIMO) systems with a large number of antennas 
have attracted a lot of attention \cite{lmimo1}-\cite{scale}. 
Such large MIMO systems are attractive because of their high spectral and power 
efficiencies. Large-scale multiuser MIMO systems where each base station (BS) 
is equipped with a large number of antennas and the user terminals are equipped 
with one or more antennas each are being studied widely. On the uplink (user 
terminals to BS link) in such large-scale MIMO systems, reduced complexity
receivers are essential at the BS for practical implementation. In addition, 
these receivers need to achieve good performance to ensure good power 
efficiencies. Channel estimation, signal detection, and channel decoding 
are key receiver functions of interest. In this paper, we are interested
in low complexity signal detection and channel decoding. In particular,
we propose a {\em non-binary belief propagation} approach for $M$-ary signal 
detection and $q$-ary LDPC decoding in large-scale MIMO systems. We also 
study the performance of the proposed detection-decoding scheme with estimated
channel knowledge.

Linear detectors like the minimum mean square error (MMSE) detector are good 
in terms of both complexity and performance when the number of BS antennas is 
much larger than the number of uplink users (i.e., low system loading factors)
\cite{mmse1}. Several algorithms with complexities comparable to that of (or 
even less than that of) MMSE detection have been shown to achieve near-optimal 
performance in large-scale MIMO systems 
\cite{lmimo1},\cite{lmimo2},\cite{las1}-\cite{heuris1}.

Belief propagation (BP) on graphs is an efficient approach for signal 
processing in large dimensions \cite{bp1},\cite{bp3}. 
In \cite{jstsp2}, a MIMO detection algorithm for binary modulation, based on 
approximate message passing on a factor graph is presented. We refer to this 
algorithm in \cite{jstsp2} as {\em binary-BP} (B-BP) algorithm. The total 
complexity of the B-BP algorithm is very low (quadratic in the number of 
dimensions) because of its Gaussian approximation of interference. Though the 
performance of the B-BP algorithm in large dimensions is very good for binary 
modulation (BPSK), its performance in higher-order QAM is rather poor (we will 
see this in results/discussions in Sec. \ref{sec3}). The BP algorithm in \cite{gta} 
uses a different approach. It obtains a tree that approximates the fully-connected 
MIMO graph and performs message passing on this tree. The performance of this 
detection algorithm for higher-order QAM is also far from optimal. 

Non-binary BP approach is known to achieve good performance at low complexities 
for $q$-ary LDPC codes \cite{davey}. In this paper, we propose a {\em non-binary BP 
(NB-BP) approach for both detection as well as decoding} which achieves very good 
complexity and performance in large-scale multiuser MIMO systems. Our new 
contributions in this paper can be summarized as follows: 
\vspace{-2mm}
\begin{itemize}
\item 	First, we propose a NB-BP based detection algorithm for $M$-ary 
	modulation, where ($i$) the messages passed between nodes are 
	constructed as vector messages, and ($ii$) the interference is 
	approximated as a scalar Gaussian random variable. While the scalar 
	approximation contributes to achieving very low complexity ({\em lower 
	than MMSE complexity}), the vector nature of the messages contribute 
	to achieving close to optimal performance in large dimensions. 
\item 	Next, through the EXIT curve matching, we obtain $q$-ary 
	LDPC codes that are optimized for the proposed NB-BP detector and 
	the LDPC decoder. These optimized irregular $q$-ary LDPC codes with 
	NB-BP detection outperform off-the-shelf irregular $q$-ary LDPC codes 
	with MMSE detection, by 1 to 7 dB at $10^{-5}$ coded BER for various 
	loading factors; number of users is varied from 16 to 128 
	and number of BS antennas is fixed at 128.  
\item 	Even under estimated channel knowledge, the proposed NB-BP detector 
	outperforms the MMSE detector.
\end{itemize}

To our knowledge, non-binary BP for detection of $M$-ary modulation and $q$-ary
LDPC code optimization for large-scale multiuser MIMO systems have not been 
reported so far.

\section{System Model}
\label{sec2} 
\begin{figure}
\includegraphics[width=3.25in,height=2.3in]{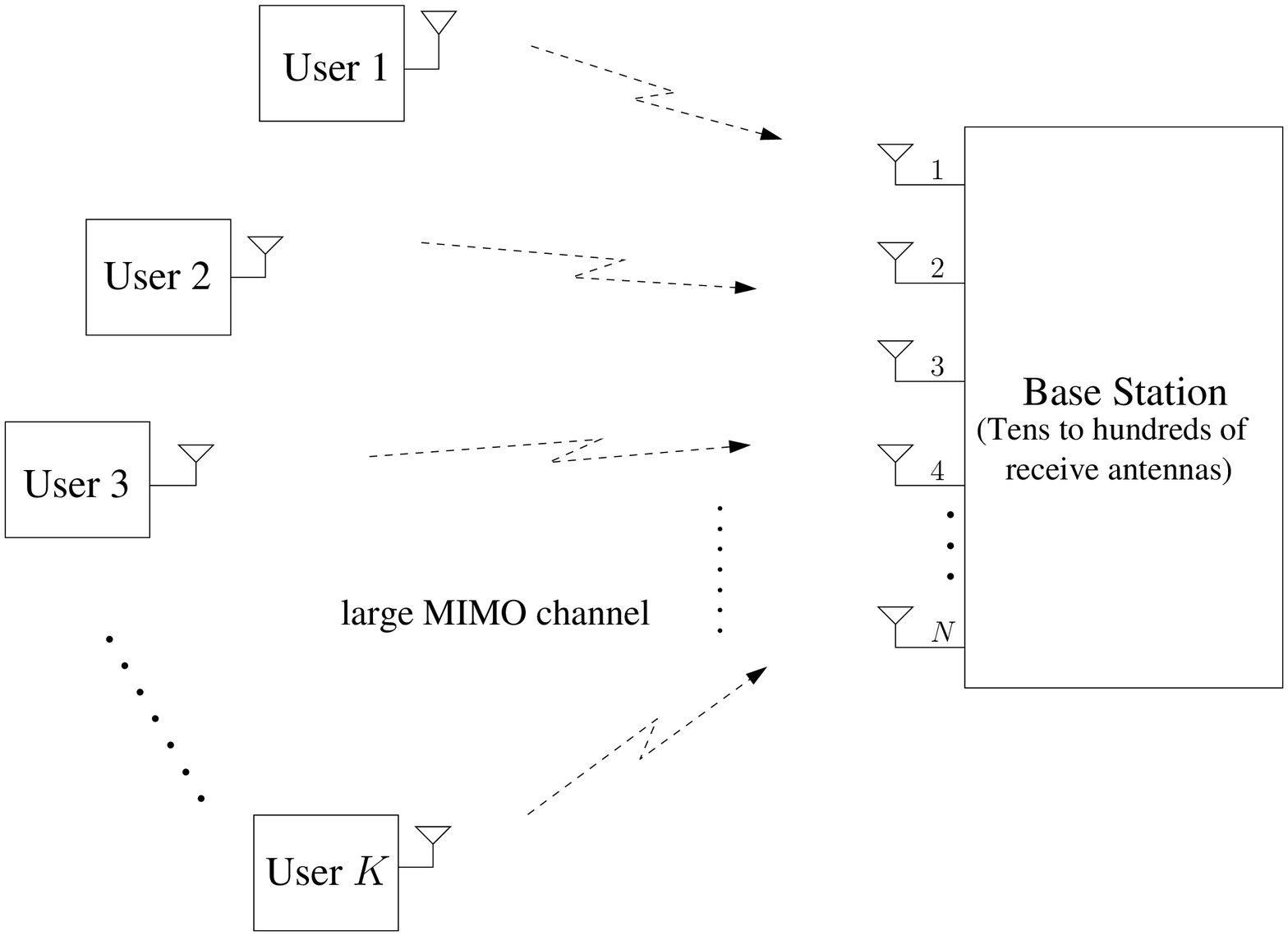} 
\caption{An illustration of the uplink multiuser-user MIMO system with $K$ 
transmitting users and $N$ receiving antennas at the base station.}
\label{system} 
\end{figure}

Consider a large-scale multiuser MIMO system where $K$ uplink users, each 
transmitting with a single antenna, communicate with a BS having a large 
number of receive antennas. Let $N$ denote the number of BS antennas; $N$ 
is in the range of tens to hundreds. The ratio $\alpha=K/N$ is the system
loading factor. This system model is illustrated in Fig. \ref{system}.
Each user uses an LDPC code over GF($q$) and $M$-QAM modulation. Each user 
encodes a sequence of $k\beta$ information bits to a sequence of $n$ coded 
symbols using a $q$-ary LDPC code with parity check matrix ${\bf F}$ 
and code rate $R=\frac{k}{n}$, where $\beta=\log_2q$. These coded symbols 
are then $M$-QAM modulated and transmitted. Assume $M=q=2^{2i}$, where $i$ 
is any positive integer. The transmission of one LDPC code block requires 
$n$ channel uses. Let ${\bf H}_c^{(t)} \in \mathbb{C}^{N\times K}$ denote 
the channel gain matrix in the $t$th channel use and $h_{ij}^c$ denote the 
complex channel gain from the $j$th user to the $i$th BS antenna. The channel 
gains $h_{ij}^c$'s are assumed to be independent Gaussian with zero mean and 
variance $\sigma_j^2$, such that $\sum_j \sigma_j^2=K$. $\sigma_k^2$ models 
the imbalance in the received power from user $k$ due to path loss etc.,
and $\sigma_j^2=1$ corresponds to the case of perfect power control. Let 
${\bf x}_c^{(t)}$ denote the modulated 
symbol vector transmitted in the $t$th channel use, where the $j$th element of 
${\bf x}_c^{(t)}$ denotes the modulation symbol transmitted by the $j$th user. 
Assuming perfect synchronization, the received vector at the BS in the $t$th 
channel use, ${\bf y}_c^{(t)}$, is given by
\begin{eqnarray}
{\bf y}_c^{(t)} & = & {\bf H}_c^{(t)}{\bf x}_c^{(t)} + {\bf w}_c^{(t)},
\label{eqn1}
\end{eqnarray}
where ${\bf w}_c^{(t)}$ is the noise vector whose entries are modeled as
i.i.d. $\mathcal{CN}(0,\sigma^2)$. Dropping the channel use index for 
convenience, (\ref{eqn1}) can be written in the real domain as
\begin{equation}
{\bf y} = {\bf Hx} + {\bf w},
\end{equation}
where
\[{\bf y} \Define
\left[\begin{array}{c} \Re({\bf y}_c) \\ \Im({\bf y}_c) \end{array}\right], \,
{\bf H} \Define \left[\begin{array}{cc}\Re({\bf H}_c) \hspace{2mm} -\Im({\bf H}_c) \\
\Im({\bf H}_c)  \hspace{5mm} \Re({\bf H}_c) \end{array}\right],
\] 
\[{\bf x} \Define
\left[\begin{array}{c} \Re({\bf x}_c) \\ \Im({\bf x}_c) \end{array}\right], \, 
{\bf w} \Define
\left[\begin{array}{c} \Re({\bf w}_c) \\ \Im({\bf w}_c) \end{array}\right],
\]
and $\Re(.)$ and $\Im(.)$ denote the real and imaginary parts, respectively.
Note that for $M$-QAM modulation, the elements of ${\bf x}$ come from the
underlying PAM alphabet $\mathbb{A} = \{\pm 1, \pm 3, \cdots, \pm \sqrt{M}-1\}$.
The BS observes ${\bf y}$ and performs detection and decoding. 

\section{Non-Binary BP for Detection}
\label{sec3}
In this section, we propose a NB-BP scheme for 
the detection of $M$-QAM symbols suited well for large-scale MIMO systems. 
A key component of the scheme is the proposed construction of vector messages 
over a factor graph using a scalar Gaussian approximation of interference. 
While the scalar approximation contributes to achieving very low complexity, 
the vector nature of the messages contributes to achieving very good performance.

\begin{figure}
\centering
\includegraphics[width=3.0in,height=1.5in]{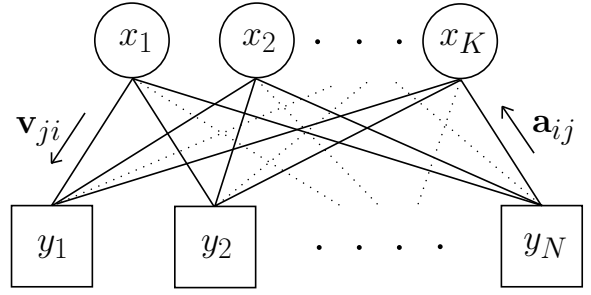} 
\caption{The factor graph and vector messages.	
\label{nbbpg} }
\vspace{-2mm}
\end{figure}

We model the system as a factor graph with $N$ observation nodes and $K$
variable nodes (see Fig. \ref{nbbpg}), and perform an approximate marginalization 
of the symbol probabilities over this graph. The $i$th element in the received 
vector ${\bf y}$ can be written as
\begin{eqnarray}
y_{i}=h_{ij}x_{j}+z_{ij}, \quad i=1,\cdots,2N, \,\, j=1,\cdots,2K,
\end{eqnarray}
where $z_{ij}\Define \sum\limits_{l=1,l\neq j}^{2K}h_{il}x_{l}+w_{i}$ is the 
interference-plus-noise term, $x_j$ is the $j$th element in ${\bf x}$, $h_{ij}$ 
is the $(i,j)$th element in ${\bf H}$, and $w_i$ is the $i$th element in ${\bf w}$. 
As in \cite{jstsp2}, we approximate the scalar term $z_{ij}$ as Gaussian 
r.v.\footnote{{\em Remark:} Although a similar scalar approximation of interference 
is used in \cite{jstsp2}, here the proposed formulation of messages as vectors 
is different and it achieves significantly improved $M$-QAM detection performance 
compared to the scalar messages based BP in \cite{jstsp2}.} with mean ($\mu_{ij}$) 
and variance ($\sigma_{ij}^2$), given by

\vspace{-2mm}
{\small
\begin{eqnarray}
\mu_{ij}=\sum_{l=1,l\neq j}^{2K}h_{il}\mathbb{E}(x_{l}), \quad 
\sigma_{ij}^{2}=\sum_{l=1,l\neq j}^{2K}h_{il}^{2}\, \text{Var}(x_{l})+\sigma^{2}.
\label{mean-eq}
\end{eqnarray}}

\vspace{-2mm}
{\em Construction of vector messages:}
Let ${\bf a}_{ij}$ denote the message passed from the $i$th observation node to
the $j$th variable node, and ${\bf v}_{ji}$ denote the message passed from the 
$j$th variable node to $i$th observation node. ${\bf a}_{ij}$ and ${\bf v}_{ji}$ 
are vectors of size $\sqrt{M}\times 1$, and they are constructed to be functions 
of the approximate likelihood and posterior probabilities. Using the Gaussian 
approximation made above, the likelihood and the posterior probabilities can be 
approximated as

\vspace{-2mm}
{\small
\begin{eqnarray}
\Pr(y_i|{\bf H},x_j=s) \approx \frac{1}{\sigma_{ij}\sqrt{2\pi}}\exp\bigg({\frac{-(y_i-\mu_{ij}-h_{ij}s)^2}{2\sigma_{ij}^2}}\bigg),
\end{eqnarray}
}

\vspace{-2mm}
where $s\in \mathbb{A}$, and
\begin{eqnarray}
\Pr(x_j=s|{\bf y},{\bf H}) & \propto & \prod_{i=1}^{2N} \Pr(y_i|{\bf H},x_j=s) \nonumber \\
& \approx & \prod_{i=1}^{2N} \frac{\exp\Big({\frac{-(y_i-\mu_{ij}-h_{ij}s)^2}{2\sigma_{ij}^2}}\Big)}{\sigma_{ij}},
\end{eqnarray}
respectively. With these approximations, the messages are defined as 

\vspace{-2mm}
{\small
\begin{eqnarray}
a_{ij}(s)&=&\frac{1}{\sigma_{ij}\sqrt{2\pi}}\exp\Big({\frac{-(y_i-\mu_{ij}-h_{ij}s)^2}{2\sigma_{ij}^2}}\Big),
\label{eq3}\\
v_{ji}(s)&=&\prod_{l=1,l\neq i}^{2N} a_{lj}(s).
\label{eq4}
\end{eqnarray}}

\vspace{-2mm}
where $a_{ij}(s)$ and $v_{ij}(s)$ are the elements of ${\bf a}_{ij}$ and 
${\bf v}_{ij}$, respectively, corresponding to the symbol $s$. The mean and 
variance at the $i$th observation node are computed as

\vspace{-2mm}
{\small
\begin{eqnarray}
\mu_{ij}&=&\sum_{l=1,l\neq j}^{2K} h_{il} {\bf s}^T{\bf v}_{li}, \label{eq1} \\
\sigma_{ij}^2&=&\sum_{l=1,l\neq j}^{2K} h_{il}^{2}\, \Big(({\bf s}\odot{\bf s})^T{\bf v}_{li}-({\bf s}^T{\bf v}_{li})^2\Big)
+\sigma^{2}. \label{eq2}
\end{eqnarray}}

\vspace{-2mm}
where ${\bf s}$ is the vector of all elements in $\mathbb{A}$ (e.g., for $M=16$, 
${\bf s}=[-3\hspace{1mm}-1\hspace{1mm}+1\hspace{1mm}+3]^T$), and $\odot$ denotes
the Hadamard product of vectors.

{\em Message passing:}\\
{\bfseries {\em Step 1)}} Initialize the posterior probability values 
$v_{ji}(s)$'s as $1/\sqrt{M}$. \\
{\bfseries {\em Step 2)}} Compute 
${\bf a}_{ij}$ messages using (\ref{eq1}), (\ref{eq2}), and (\ref{eq3}). \\
{\bfseries {\em Step 3)}} Compute 
${\bf v}_{ji}$ messages using (\ref{eq4}). This forms one iteration of the 
algorithm. \\
Repeat Steps 2) and 3) for a certain number of iterations. Damping on 
${\bf v}_{ji}$ messages can be done in the $m$th iteration using a 
damping factor $\delta$ as
\begin{eqnarray}
{\bf v}_{ji}^{(m)}=(1-\delta){\bf v}_{ji}^{(m)} + \delta {\bf v}_{ji}^{(m-1)},
\quad \delta \in [0,1).
\end{eqnarray}
After a given number of iterations, the final symbol probabilities are computed as
\begin{eqnarray}
P_{x_j}(s) \ \Define \ \Pr(x_j=s) \ \propto \ \prod_{l=1}^{2N} a_{lj}(s).
\end{eqnarray}
A listing of the proposed NB-BP algorithm is listed in {\bf Algorithm 1}.
The $P_{x_j}(s)$ values $\forall s, j$ are fed as soft inputs to the $q$-ary LDPC 
decoder. In uncoded systems, hard bit decisions are made on the bit probability 
values computed as 
\begin{eqnarray}
\Pr(b_j^p=1)=\sum_{\forall s\in\mathbb{A} : \ p\mbox{\scriptsize{th bit in}} \hspace{0.5mm} s\hspace{0.5mm} \mbox{\scriptsize{is}} \hspace{0.5mm} 1}P_{x_j}(s),
\end{eqnarray}
where $b_j^p$ is the $p$th bit in the $j$th user's symbol. 

{\em Complexity:} From (\ref{eq4}), (\ref{eq1}) and (\ref{eq2}), the total 
complexity of the NB-BP detector proposed above is $O(KN\sqrt{M})$. 
This is because the summations in (\ref{eq1}), (\ref{eq2}) can be 
computed by summing over all node indices once and subtracting from it the term 
to be excluded at each node. 
{\em Note that the $O(KN\sqrt{M})$ complexity of the proposed detector is 
less compared to the MMSE detector complexity of $O(KN^2)$ for large $N$
and moderately sized QAM. This is because 
the MMSE detector needs a matrix inversion, whereas the proposed NB-BP detector 
does not need a matrix inversion. More interestingly, as discussed next, even 
with this less than MMSE complexity, the proposed NB-BP detector performs 
increasingly closer to optimal performance in large-scale MIMO systems.} 

\vspace{4mm}
\hrule

\begin{algorithm}
\vspace{-2mm}
{\small
\SetLine
\KwIn{${\bf y}$, ${\bf H}$, $\sigma^2$}
{\bf Initialize}: $v_{ji}^{(0)}(s)\gets1/{\sqrt M}$, $\forall i,j,s$

\For{$m = 1 \to {\textit Number\_of\_iterations}$ }{
\For{$i = 1 \to 2N$ }{
$\mu_{i}\gets\sum\limits_{l=1}^{2K} h_{il} {\bf s}^T{\bf v}_{li}^{(m-1)}$

$\sigma_{i}^2\gets\scriptstyle{\sum\limits_{l=1}^{2K} h_{il}^{2}\, \Big(({\bf s}\odot{\bf s})^T{\bf v}_{li}^{(m-1)}-({\bf s}^T{\bf v}_{li}^{(m-1)})^2\Big)+\sigma^{2}}$

\For{$j = 1 \to 2K$ }{
$\mu_{ij}\gets\mu_{i}-h_{ij}{\bf e}^T{\bf v}_{ji}^{(m-1)}$
$\sigma_{ij}^2\gets\scriptstyle{\sigma_{i}^2-h_{ij}^{2}\, \Big(({\bf s}\odot{\bf s})^T{\bf v}_{ji}^{(m-1)}-({\bf s}^T{\bf v}_{ji}^{(m-1)})^2\Big)+\sigma^{2}}$

\ForEach{$s \in \mathbb{A}$}{
$a_{ij}(s)\gets\frac{1}{\sigma_{ij}\sqrt{2\pi}}\exp\Big({\frac{-(y_i-\mu_{ij}-h_{ij}s)^2}{2\sigma_{ij}^2}}\Big)$
}
}
}

\For{$j = 1 \to 2K$ }{
\ForEach{$s \in \mathbb{A}$}{
$v_{j}^{(m)}(s)\gets\prod\limits_{l=1}^{2N} a_{lj}(s)$
}
\For{$i = 1 \to 2N$ }{
\ForEach{$s \in \mathbb{A}$}{
$v_{ji}^{(m)}(s)\gets v_{j}(s)/a_{ij}(s)$
}
${\bf v}_{ji}^{(m)}=(1-\delta){\bf v}_{ji}^{(m)} + \delta {\bf v}_{ji}^{(m-1)}$
}
}
}
\KwOut{$P_{x_j}(s)\gets \frac{1}{Z}\prod\limits_{l=1}^{2N} a_{lj}(s)$,
$Z$ is normalizing constant.}
\label{nbbp}

\vspace{2mm}
\caption{The proposed NB-BP detection algorithm.}
}
\end{algorithm}
\vspace{-2mm}
\hrule
\vspace{4mm}

{\em Reducing computational complexity:} The computational complexity of the
proposed NB-BP scheme is dictated by the computations required for the terms
$\mu_i$ and $\sigma_i^2$. The number of operations required for computing 
$\mu_i$ in the form presented in (\ref{eq1}) is $(1+2\sqrt M)2K-1$. By 
distributive law, the terms can be rearranged as
\begin{equation}
\mu_i=\sum_{\forall s}s\sum_{l=1}^{2K}v_{li}(s)h_{il}.
\end{equation}
This gives a lesser computational complexity of $(1+4K){\sqrt M}-1$ (for 
$\sqrt M < 2K$). Further, it can be noted that the double summation involved 
in the computation of $\mu_i$ and $\sigma^2_i$ can be viewed as a convolution 
operation, and hence the complexity can be further reduced by using FFT
to compute the convolutions.

\begin{figure}
\hspace{-5mm}
\includegraphics[width=3.75in,height=2.65in]{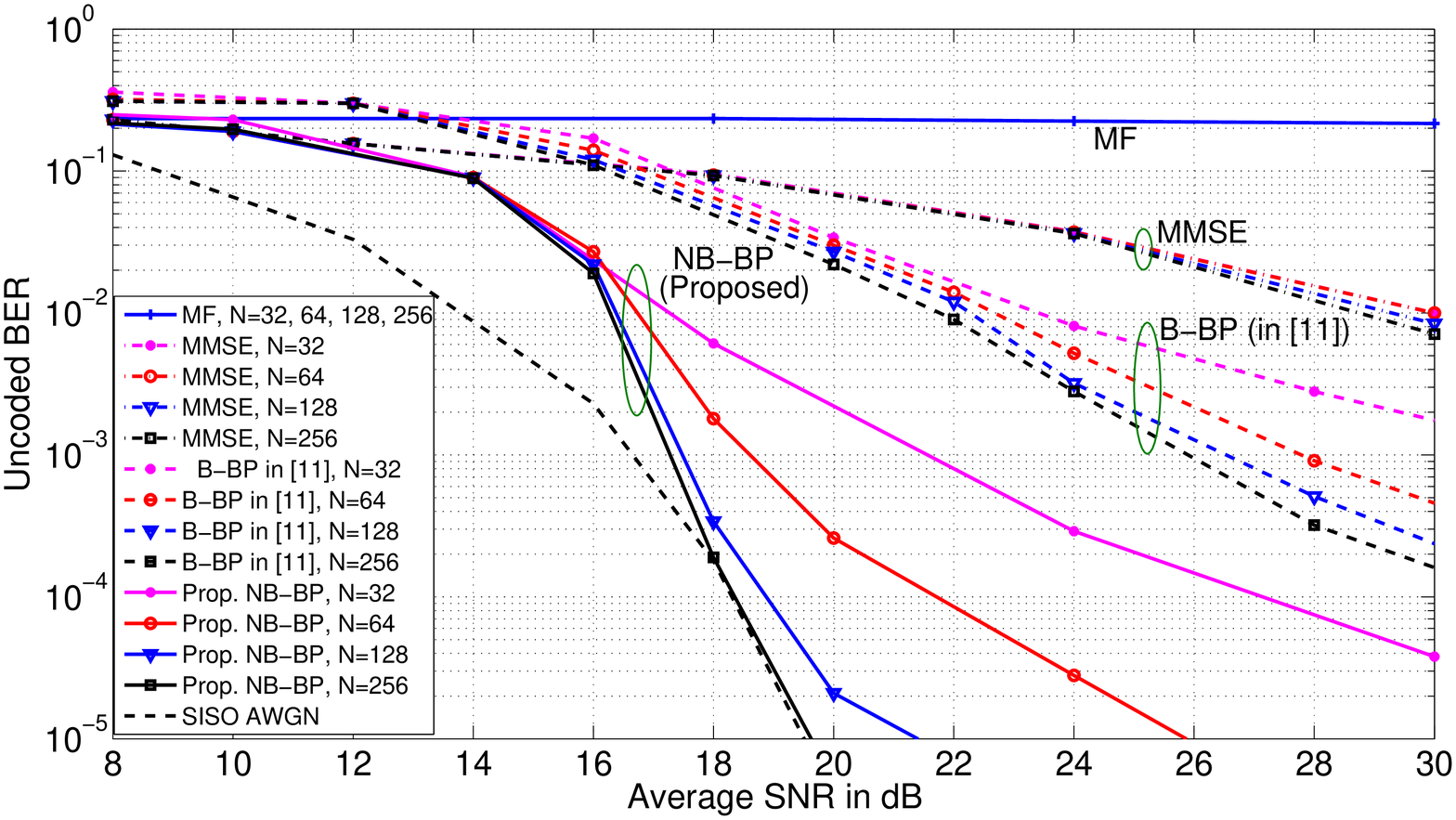} 
\caption{Comparison of uncoded BER performance of the proposed NB-BP detector 
with those of B-BP in \cite{jstsp2}, MMSE and MF detectors for $N=32,64,128,256$, 
$\alpha=1$, and 16-QAM.}
\label{detp2p} 
\end{figure}

{\em BER Performance:}
Figure \ref{detp2p} shows the uncoded BER performance of the proposed NB-BP detector 
for 16-QAM in multiuser MIMO with $N=32,64,128,256$, $\alpha=1$, and $\sigma_j=1$. 
The performance of the B-BP detector in \cite{jstsp2}, MMSE detector, 
MF detector, and unfaded SISO AWGN performance are also plotted for comparison. 
For using the B-BP scheme in \cite{jstsp2} for $M$-QAM detection, each $M$-QAM 
symbol is written in the form of linear combination of the constituent $q$ bits 
and the equivalent system model is written as 
${\bf y}={\bf H}({\bf I}_K\otimes{\bf m}){\bf x}_b +{\bf w}$, where 
${\bf m}=[2^0 \ 2^1 \ \cdots \ 2^{\frac{\beta}{2}-1}]$, ${\bf x}_b\in\{\pm1\}^{K\beta}$ 
is the vector of information bits, and the B-BP algorithm is run on the equivalent 
bit-level system model with the resulting complexity being the same
as that of NB-BP. In the simulations, the number of BP iterations used is 40 and 
the damping factor used is $\delta=0.2$. From Fig. \ref{detp2p}, we observe that the 
NB-BP detector performs considerably better than the MMSE and MF detectors. 
In large dimensions (e.g., $N=256$), the NB-BP detector performance gets very 
close to SISO-AWGN performance. Also, the NB-BP scheme significantly outperforms 
the B-BP scheme (e.g., for $N=256$, NB-BP performs better than B-BP by about 8 dB 
at $10^{-3}$ BER). This is because, with $M$-QAM, the assumption that the elements 
of ${\bf x}_b$ in B-BP are independent is not true, and this results in a degraded 
performance in B-BP when applied to $M$-QAM. 

\begin{figure}
\centering 
\hspace{-1mm}
\subfigure[Performance]{
\includegraphics[width=1.675in,height=2.25in]{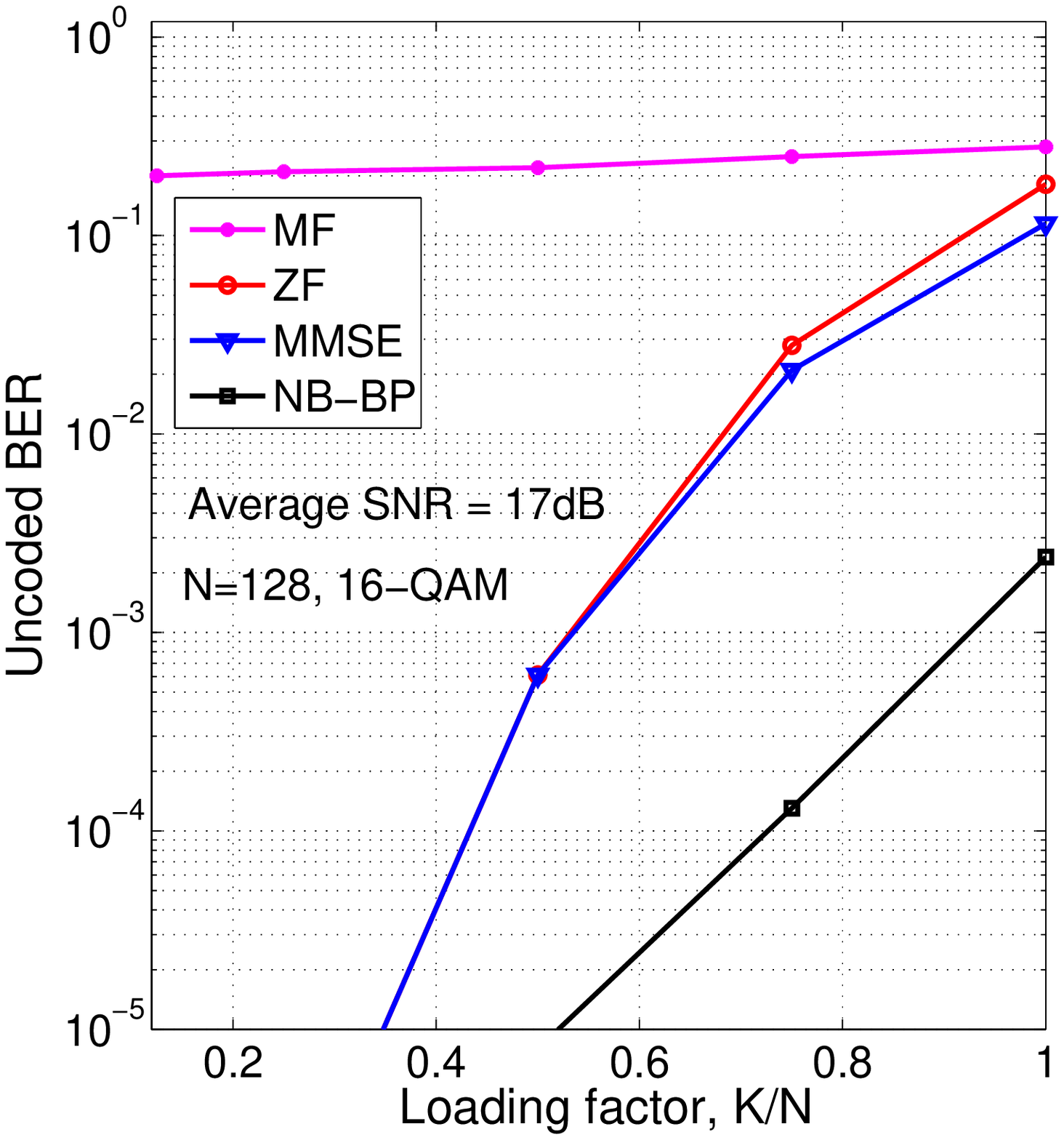}
\label{detmu}
}
\hspace{-6mm}
\subfigure[Complexity]{
\includegraphics[width=1.675in,height=2.25in]{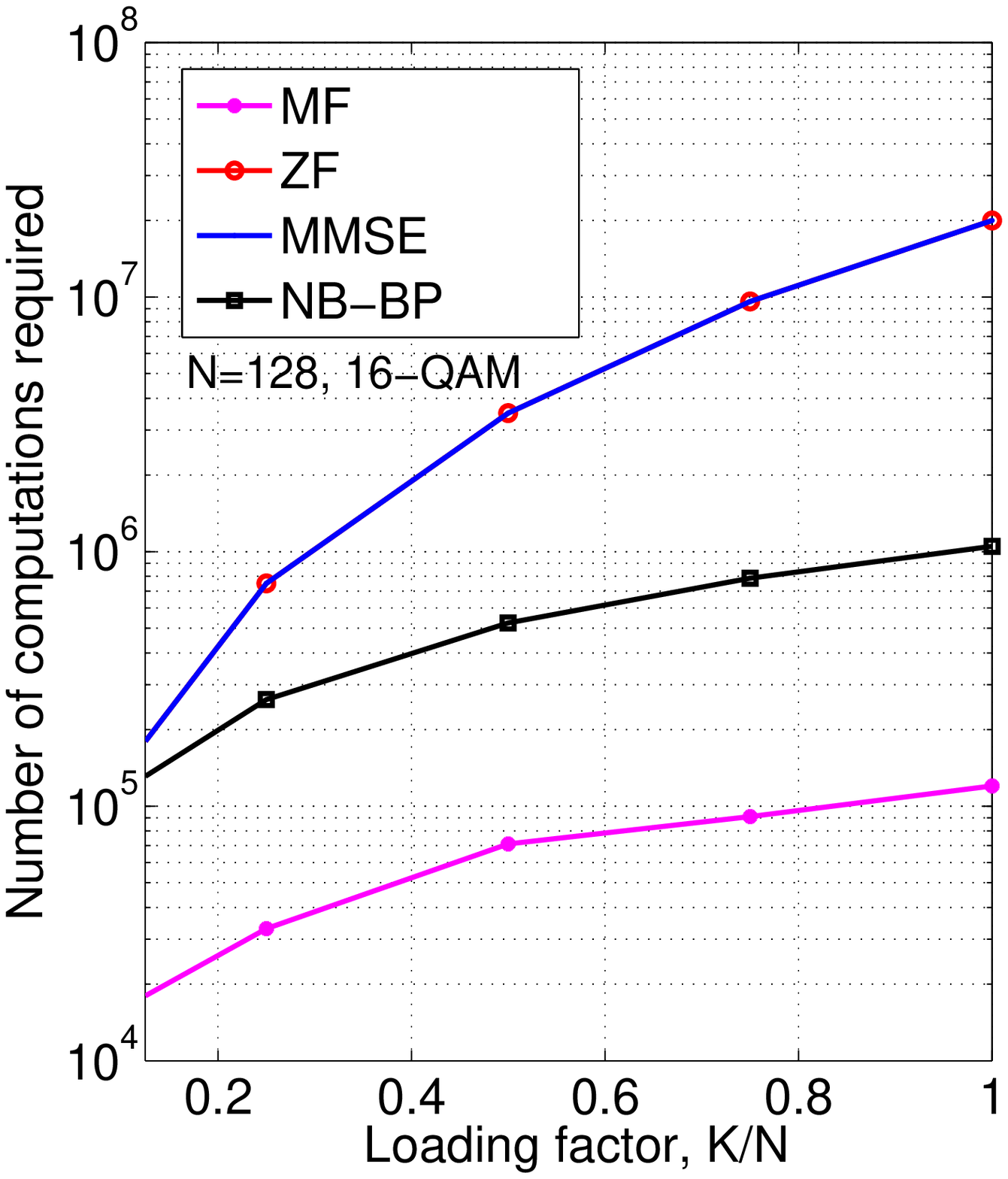}
\label{cdetmu}
}
\caption{Comparison of uncoded BER performance and complexity of the proposed 
NB-BP detector with those of linear detectors (MMSE, ZF, MF) as a function of 
loading factor $\alpha$, with $N=128$ and 16-QAM.}
\end{figure}

Next, in Figs. \ref{detmu} and \ref{cdetmu}, we show performance and complexity 
comparison of NB-BP with MMSE, ZF, and MF detectors for varying loading factors at 
17 dB SNR, $N=128$ and 16-QAM. As mentioned earlier, we can see that the NB-BP 
scheme achieves better performance than ZF and MMSE detectors at lesser complexity
than these detectors across various loading factors, $\alpha$.

\section{Optimized LDPC Code Design for NB-BP Detector-Decoder}
\label{sec4}
In this section, we compute the EXIT chart of the NB-BP detector combined with the 
$q$-ary LDPC decoder, and obtain the optimal degree profile distribution of the 
$q$-ary LDPC code.  The $q$-ary LDPC codewords are decoded by a message passing 
algorithm over a bipartite graph made of $n$ variable nodes and $k$ check nodes. 
A detailed description of non-binary LDPC decoding algorithm can be found in 
\cite{davey},\cite{bennatan}. As in \cite{vtc}, we formulate an integrated 
graph consisting of three sets of nodes, namely, variable nodes set, observation 
nodes set, and check nodes set. There are $nN$ observation nodes corresponding 
to the received vectors, $nK$ variable nodes corresponding to the transmitted 
coded symbol vectors, and $K(n-k)$ check nodes corresponding to the check 
equations of the LDPC code. Combining the NB-BP detector proposed in the previous 
section and the non-binary LDPC decoder, a joint message passing scheme is 
formulated for joint detection-decoding. 

\begin{table*}
\centering
\begin{tabular}{|c||p{4.4cm}|p{4.4cm}|p{4.4cm}|}
\hline
Parameters & $N=128$, $\alpha=1$ & $N=128$, $\alpha=0.5$ & $N=128$,
$\alpha=0.25$
\tabularnewline
\hline \hline
 ($d_v$, $p_v$)& (2, 0.4768), (6, 0.0104), (8, 0.3174), \newline
 (12, 0.1817), (16, 0.0024), (20, 0.0113) &
 (2, 0.6246), (8, 0.168) (16, 0.1853),\newline
 (20, 0.0221) &
 (2, 0.3557), (3, 0.6018), (8, 0.0067),\newline
 (12, 0.0358) \tabularnewline
\hline
 ($d_c$, $p_c$)& (6, 0.5206), (10, 0.1973), (18, 0.1517), \newline
 (32, 0.1304) &
 (8, 0.5649), (16, 0.1755), (18, 0.2596) &
 (5, 0.7287), (8, 0.1793), (10,        0.0922) \tabularnewline
\hline
\end{tabular}
\caption{Degree profiles of optimized rate-1/2 16-ary LDPC codes for different 
large multiuser MIMO configurations. 
$p_v$, $p_c$: fraction of variable nodes with degree $d_v$ and check nodes with
degree $d_c$. 
}
\label{tab1}
\vspace{-3mm}
\end{table*}

{\em EXIT analysis:}
We use the EXIT chart analysis for analyzing the behavior of joint detector-decoder. 
If $I_A$ is the average mutual information between the coded symbols and input 
a priori information, and $I_E$ is the average mutual information between the 
coded symbols and the extrinsic output, then the EXIT function is $f(I_A)=I_E$. 
To obtain the EXIT characteristics of the joint detector-decoder, we first obtain 
the EXIT curves of the NB-BP detector and combine it with that of the LDPC decoder.

Let $I_{E,nbbp}$ and $I_{A,nbbp}$ denote the $I_E$ and $I_A$, respectively,
for the NB-BP detector. Then the EXIT function is 
$I_{E,nbbp}=f\left(\gamma, K, N, I_{A,nbbp}\right)$, where $\gamma$ is the average 
received SNR. Since
an analytical evaluation of this function is difficult, we compute it through 
Monte Carlo simulations \cite{tenbrink1}. If $d_v$ and $d_c$ denote the variable 
node and check node degrees, respectively, of the LDPC code, then the EXIT 
function \cite{bennatan}, \cite{tenbrink1} of the LDPC variable nodes set 
is given by $I_{E,V}=J\left(\sqrt{(d_v-1)(J^{-1}(I_{A,V}))^2+c\gamma}\right)$, 
and the EXIT function of the LDPC check nodes set is given by
$I_{E,C}=1-J\left(J^{-1}(1-I_{A,C})\sqrt{(d_c-1)}\right)$, where $c$ is 
a constant dependent on $q$ and $J(.)$ is as defined in \cite{tenbrink1}.

We formulate the EXIT function of the combination of the NB-BP detector and the
variable nodes set of the $q$-ary LDPC decoder as

\vspace{-2mm}
{\small
\begin{eqnarray}
I_{E,JV}(\gamma, d_v, I_{A,JV}, I_{E,nbbp})= \hspace{30mm}\nonumber \\
J\left(\sqrt{(d_v-1)(J^{-1}(I_{A,JV}))^2+(J^{-1}(I_{E,nbbp}))^2}\right),
\label{combined}
\end{eqnarray}
}

\vspace{-2mm}
where $I_{E,JV}$  and $I_{A,JV}$ are the $I_E$ and $I_A$, respectively, for the
combined variable nodes set. We match this EXIT curve with that of the check nodes
set, such that the EXIT curve of the check nodes set lie below the EXIT curve of 
the combined variable nodes set.

\begin{figure}
\hspace{-4mm}
\includegraphics[width=3.75in,height=2.65in]{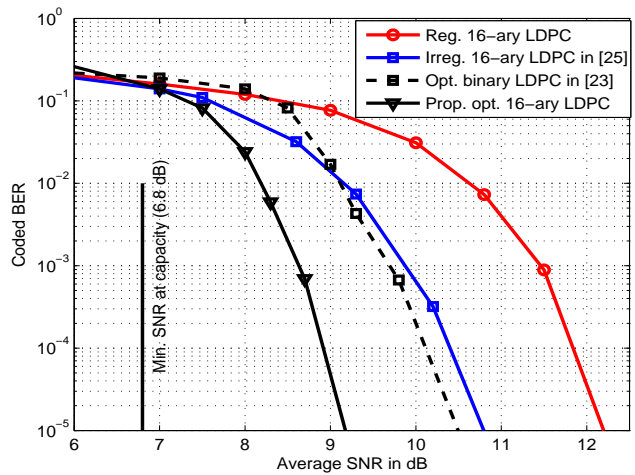}
\vspace{-4mm}
\caption{Performance comparison of the proposed irregular non-binary LDPC
codes with other LDPC codes for $n=1000$ coded symbols, 16-QAM and $N=K=64$, 
at a spectral efficiency of 128 bits/s/Hz.
\label{ldpc}}
\vspace{-2mm}
\end{figure}

{\em Optimized LDPC code design procedure:}
To approach the capacity of the channel, the EXIT curves of the check nodes
set and the variable nodes set should be matched. This matching 
is done by obtaining an appropriate degree distribution of the variable nodes
and the check nodes, thereby designing irregular LDPC codes for a specific 
channel and receiver. The design methodology we adopt is described 
in \cite{comm}. We use the method described in \cite{symb_ent} to optimize the 
non-zero entries of the parity check matrix. By this method, the combination of 
the non-zero entries of a row of the parity check matrix that maximize the average 
entropy of the syndrome vector is chosen to be the entries of the row of our parity 
check matrix, $\bf F$. We obtained optimized non-binary irregular LDPC codes using 
the design procedure described above, and the obtained codes for different system 
settings and loading factors are given in Table \ref{tab1}.

\begin{figure}
\includegraphics[width=3.5in,height=2.65in]{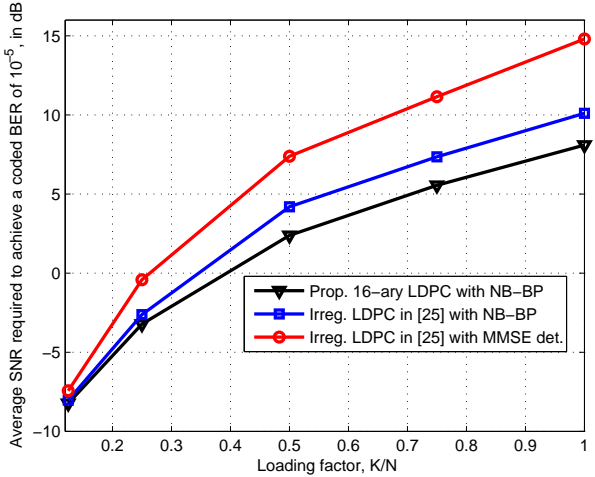}
\vspace{-4mm}
\caption{Performance comparison of the proposed irregular non-binary LDPC
codes with other LDPC codes for $n=1000$ coded symbols and 16-QAM. 
$N=128$ for different loading factors.
\label{ldpcmu}}
\vspace{-2mm}
\end{figure}

{\em Coded BER performance:}
We evaluated the coded BER performance of the proposed non-binary LDPC codes and
compared with those of other LDPC codes, namely, ($i$) random non-binary 
`regular' LDPC code, ($ii$) non-binary irregular LDPC code in \cite{irnb}, and 
($iii$) optimized `binary' irregular LDPC code in \cite{comm}. Figure \ref{ldpc} 
shows the simulated coded BER performance of the proposed rate-1/2 non-binary 
LDPC code with $n=1000$ coded symbols using NB-BP detection and decoding in a 
system with $N=64$, $\alpha=1$, and 16-QAM. It can be seen that the proposed code
significantly outperforms other codes; e.g., by about 1.2 to 3 dB at $10^{-5}$ 
coded BER. The better performance of the proposed code is because of the matching 
of EXIT charts of the combined NB-BP detector and non-binary LDPC decoder. Also, 
the proposed code's performance is just about 2.3 dB away from capacity. Figure 
\ref{ldpcmu} shows the average SNR required to achieve a coded BER of $10^{-5}$ 
by the proposed rate-1/2 non-binary LDPC codes with NB-BP detection as a function 
of loading factor for $N=128$, $n=1000$ coded symbols, and 16-QAM. It can be seen 
that this performance is better than the performance achieved by the non-binary
LDPC code in \cite{irnb} with MMSE detection and NB-BP detection, by 1 dB to 7 dB 
for various system loading factors. 

\section{NB-BP Detection Performance with Estimated Channel}
\label{sec5}
\begin{figure}
\hspace{1mm}
\includegraphics[width=3.25in,height=2in]{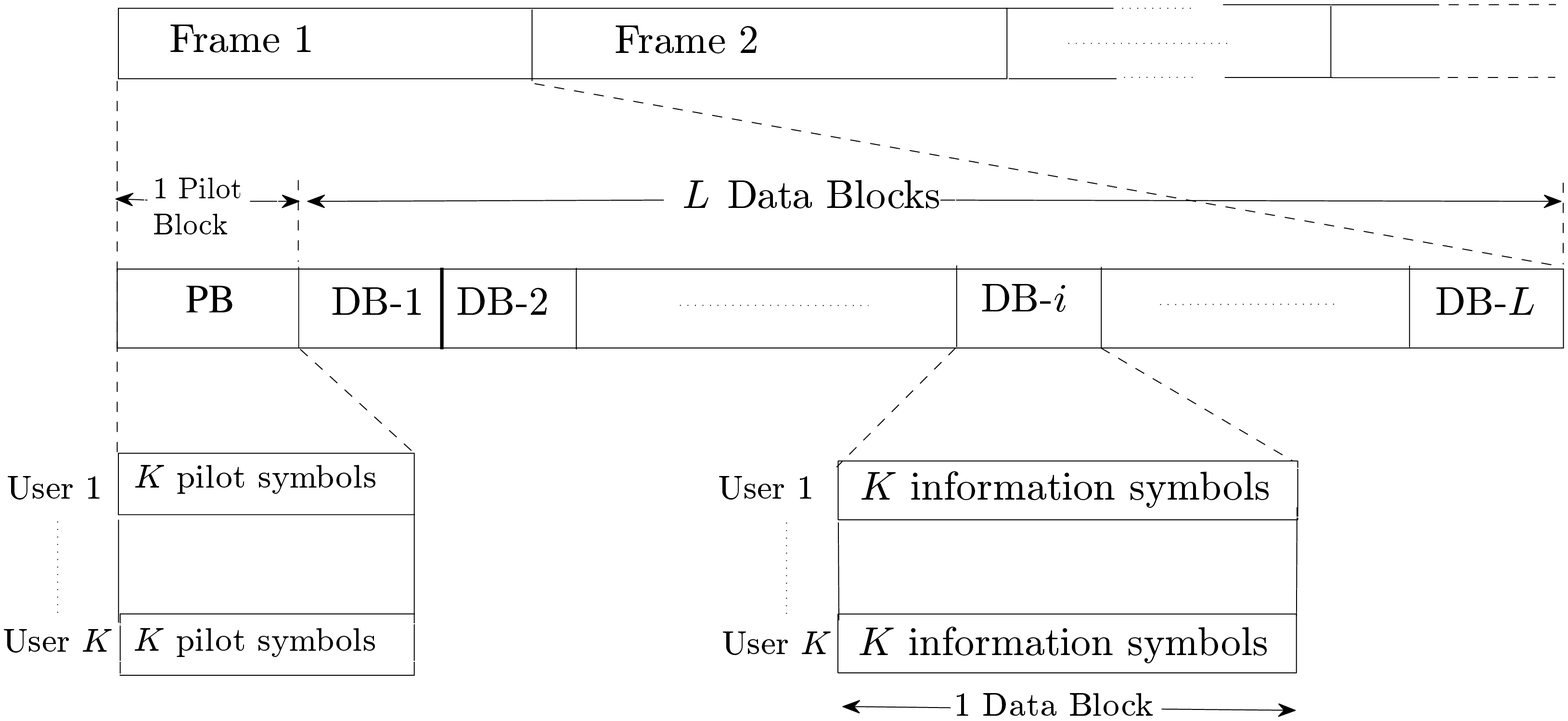}
\caption{Frame structure on the multiuser MIMO uplink.}
\label{fig_chl}
\vspace{-2mm}
\end{figure}

In evaluating the BER performance in the previous two sections, we assumed 
perfect channel knowledge at the BS receiver. In this section, we relax the 
perfect channel knowledge assumption and study the performance of the 
proposed NB-BP detector with estimated channel knowledge. Specifically,
the channel matrix is estimated based on a pilot-based channel estimation 
scheme. It is assumed that transmission is carried out in frames. It is 
further assumed that the channel remains constant over one frame duration. 
Each frame consists of a pilot block (PB) for the purpose of initial channel 
estimation, followed by $L$ data blocks (DBs) as shown in Fig. \ref{fig_chl}. 
The PB consists of $K$
channel uses in which a $K$-length pilot symbol vector comprising of pilot 
symbols transmitted from $K$ users (one pilot symbol per user) is received by
$N$ receive antennas at the BS. Each DB consists of $K$ channel uses, where
$K$ number of $K$-length information symbol vectors (one data symbol from 
each user) are transmitted. Taking both pilot and data channel uses into 
account, the total number of channel uses per frame is $(L+1)K$. 

An MMSE estimate of the channel is first obtained during the PB.
Using this estimated channel, the DBs are detected using any one of the 
detection algorithms (e.g., proposed NB-BP detector or MMSE detector). 
The detected DBs are used to refine the channel estimates during the 
data phase. The refined channel estimates are then used to again detect
the DBs, and this iteration between channel estimation and data detection
is carried out for a certain number of iterations.

\begin{figure}
\hspace{-4mm}
\includegraphics[width=3.75in,height=2.65in]{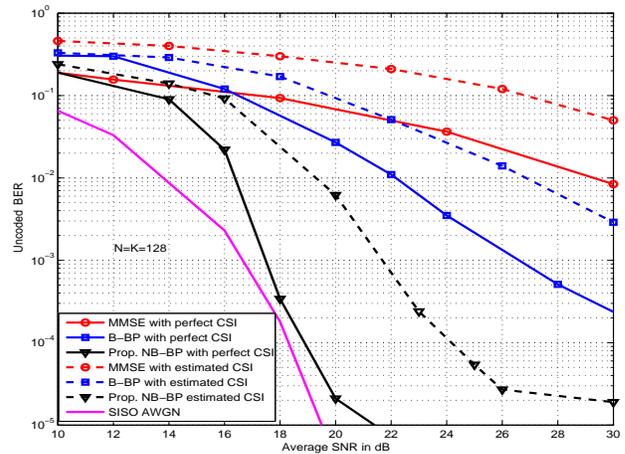}
\caption{Performance comparison between NB-BP, B-BP and MMSE detectors
with estimated CSI in large MIMO system with $N=K=128$, 16-QAM.}
\label{chl128}
\vspace{-2mm}
\end{figure}

\begin{figure}
\hspace{-4mm}
\includegraphics[width=3.75in,height=2.65in]{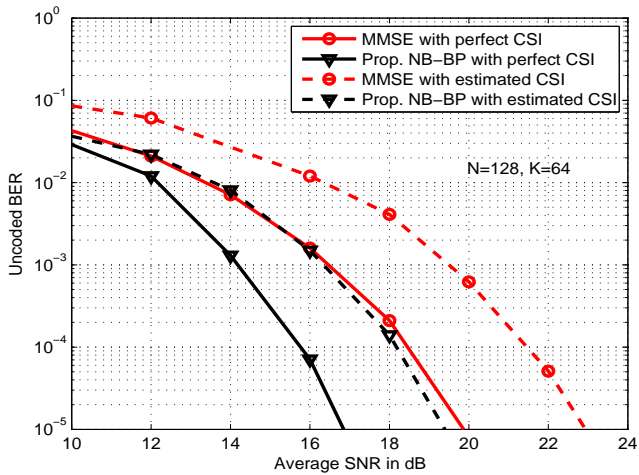}
\caption{Performance comparison between NB-BP and MMSE detectors with 
estimated CSI in large MIMO system with $N=128$, $\alpha=0.5$, 16-QAM.}
\label{chl64}
\vspace{-2mm}
\end{figure}

{\em Performance:}
Figure \ref{chl128} shows the uncoded BER performance of the proposed
NB-BP detector with estimated channel state information (CSI) for $N=K=128$, 
16-QAM, $L=4$. MMSE 
channel estimate is used. Two iterations between channel estimation and 
data detection are performed. From Fig. \ref{chl128}, we see that the 
proposed NB-BP detector performs significantly better than the B-BP and 
MMSE detectors. A similar performance advantage can be observed in 
Fig. \ref{chl64} for $N=128$ and $K=64$.

\section{Conclusions}
\label{sec6}
We proposed a promising non-binary BP algorithm for $M$-QAM signal 
detection in large-scale MIMO systems. An interesting feature of the
proposed algorithm from an implementation view point is that it 
achieves close to optimum performance in large-scale MIMO systems 
with less than MMSE complexity for large number of BS antennas 
and moderately sized QAM (which are typical in large-scale MIMO
systems). It also enabled 
the design of good LDPC codes matched to large MIMO channels.

\bibliographystyle{ieeetr}

\begin{thebibliography}{99}
\vspace{-0.0mm}
\bibitem{lmimo1}
K. V. Vardhan, S. K. Mohammed, A. Chockalingam, and B. S. Rajan,
``A low-complexity detector for large MIMO systems and multicarrier CDMA
systems,'' {\em IEEE J. Sel. Areas Commun.,} vol. 26, no. 3, pp. 473-485,
Apr. 2008.

\vspace{-0.0mm}
\bibitem{lmimo2}
S. K. Mohammed, A. Chockalingam, and B. S. Rajan, ``A low-Complexity
precoder for large multiuser MISO systems,'' {\em Proc. IEEE VTC'2008},
pp. 797-801, May 2008.

\vspace{-0.0mm}
\bibitem{lmimo3}
S. K. Mohammed, A. Zaki, A. Chockalingam, and B. S. Rajan,
``High-rate space–time coded large-MIMO systems: low-complexity detection
and channel estimation,'' {\em IEEE J. Sel. Topics Signal Proc.}, vol. 3,
no. 6, pp. 958-974, Dec. 2009.

\vspace{-0.0mm}
\bibitem{scale}
F. Rusek, D. Persson, B. K. Lau, E. G. Larsson, T. L. Marzetta, O. Edfors, and 
F. Tufvesson, ``Scaling up MIMO: opportunities and challenges with very large 
arrays,'' {\em IEEE Signal Process. Mag.}, vol. 30, no. 1, pp. 40-60, Jan. 2013.

\vspace{-0mm}
\bibitem{mmse1}
J. Hoydis, S. ten Brink, and M. Debbah, ``Massive MIMO in the UL/DL of cellular 
networks: how many antennas do we need?'' {\em IEEE J. Sel. Areas in Commun.},
vol. 31, no. 2, pp. 160-171, Feb. 2013.   

\vspace{-0mm}
\bibitem{las1}
B. Cerato and E. Viterbo, ``Hardware implementation of a low-complexity
detector for large MIMO,'' {\em Proc. IEEE ISCAS'2009}, pp. 593-596,
May 2009.

\vspace{-0mm}
\bibitem{las2}
P. Li and R. D. Murch, ``Multiple output selection-LAS algorithm in large MIMO
systems,'' {\em IEEE Commun. Lett.}, vol. 14, no. 5, pp. 399-401, May 2010.

\vspace{-0mm}
\bibitem{rts1}
N. Srinidhi, T. Datta, A. Chockalingam, and B. S. Rajan, ``Layered tabu search
algorithm for large-MIMO detection and a lower bound on ML performance,''
{\em IEEE Trans. Commun.}, vol. 59, no. 11, pp. 2955-2963, Nov. 2011.

\vspace{-0mm}
\bibitem{rts2}
T. Datta, N. Srinidhi, A. Chockalingam, and B. S. Rajan, ``Random-restart
reactive tabu search algorithm for detection in large-MIMO systems,''
{\em IEEE Commun. Lett.}, vol. 14, no. 12, pp. 1107-1109, Dec. 2010.

\vspace{-0mm}
\bibitem{hoeher}
C. Knievel, M. Noemm, and P. A. Hoeher, ``Low complexity receiver
for large-MIMO space time coded systems,'' {\em Proc. IEEE VTC'2011-Fall},
pp. 1-5, Sep. 2011.

\vspace{-0mm}
\bibitem{jstsp2}
P. Som, T. Datta, N. Srinidhi, A. Chockalingam, and B. S. Rajan, ``Low-complexity
detection in large-dimension MIMO-ISI channels using graphical Models,''
{\em IEEE J. Sel. Topics in Signal Process.}, vol. 5, no. 8, pp. 1497-1511, 
Dec. 2011.

\vspace{-0mm}
\bibitem{gta}
J. Goldberger and A. Leshem, ``MIMO detection for high-order QAM based on a
Gaussian tree approximation,'' {\em IEEE Trans. Inform. Theory},
vol. 57. no. 8, pp. 4973-4982, Aug. 2011.

\vspace{-0mm}
\bibitem{lattice1}
Q. Zhou and X. Ma, ``Element-based lattice reduction algorithms for large MIMO
detection,'' {\em IEEE J. Sel. Areas Commun.}, vol. 31, no. 2, 274-286,
Feb. 2013.

\vspace{-0mm}
\bibitem{lattice2}
K. A. Singhal, T. Datta, and A. Chockalingam, ``Lattice reduction aided detection
in large-MIMO systems,'' {\em Proc. IEEE SPAWC'2013}, pp. 589-593, Jun. 2013.

\vspace{-0mm}
\bibitem{mcmc1}
T. Datta, N. A. Kumar, A. Chockalingam, and B. S. Rajan, ``A novel
Monte-Carlo-sampling-based receiver for large-scale uplink multiuser
MIMO systems,'' {\em IEEE Trans. Veh. Tech.,} vol. 62, no. 7, pp. 3019-3038,
Sep. 2013.

\vspace{-0mm}
\bibitem{heuris1}
P. Svac, F. Meyer, E. Riegler, and F. Hlawatsch, ``Soft-heuristic detectors
for large MIMO systems,'' {\em IEEE Trans. Signal Proc.} vol. 61, no. 18,
4573-4586, Sep. 2013.

\vspace{-0mm}
\bibitem{bp1}
R. J. McEliece, D. J. C. MacKay, and J-F. Cheng, ``Turbo decoding as an instance of
Pearl’s ``belief propagation'' algorithm,'' {\em IEEE J. Sel. Areas in Commun.}, 
vol. 16, no. 2, pp. 140-152, Feb. 1998.
 
\vspace{-0mm}
\bibitem{bp3}
B. M. Kurkoski, P. H. Siegel, and J. K. Wolf, ``Joint message-passing
decoding of LDPC codes and partial-response channels,'' {\em IEEE Trans. 
Inform. Theory}, vol. 48, no. 6, pp. 1410-1422, Jun. 2002.

\bibitem{davey}
M. Davey and D. MacKay, ``Low density parity check codes over GF($q$),''
{\em IEEE Commun. Letters}, vol. 2, no. 6, pp. 165-167, Jun. 1998.

\bibitem{bennatan}
A. Bennatan and D. Burshtein. ``Design and analysis of nonbinary LDPC codes for 
arbitrary discrete-memoryless channels,'' {\em  IEEE Trans. Inform. Theory}, 
vol. 52, no.2, pp. 549-583, Feb. 2006.

\bibitem{vtc}
T. L. Narasimhan, A. Chockalingam, and B. S. Rajan, ``Factor graph based 
joint detection/decoding for LDPC coded large MIMO systems,'' {\em Proc. IEEE 
VTC'2012-Spring}, May 2012. 

\bibitem{tenbrink1}
S. ten Brink, G. Kramer, and A. Ashikhmin, ``Design of low-density parity-check
codes for modulation and detection,'' {\em IEEE Trans. Commun.,} vol. 52, no. 4,
pp. 670-678, Apr. 2004.

\bibitem{comm}
T. L. Narasimhan and A. Chockalingam, ``EXIT chart based design of irregular
LDPC codes for large MIMO systems,'' {\em IEEE Commun. Letters}, vol. 17, no. 1, 
pp. 115-118, Jan. 2013. 

\bibitem{symb_ent}
D. J. C. MacKay, ``Optimizing sparse graph codes over GF($q$),''
online: http://www.inference.phy.cam.ac.uk/mackay/CodesGallager.html

\bibitem{irnb}
A. Marinoni, P. Savazzi, and R. D. Wesel, ``Protograph-based $q$-ary LDPC codes for
higher-order modulation.,'' {\em IEEE Intl. Symp. on Turbo Codes and Iterative 
Information Processing}, Sep. 2010.

\end{thebibliography}

\end{document}